\documentclass{article}

\usepackage{arxiv}

\usepackage{bm}
\usepackage[utf8]{inputenc} 
\usepackage[T1]{fontenc}    
\usepackage{hyperref}       
\usepackage{url}            
\usepackage{booktabs}       
\usepackage{amsfonts}       
\usepackage{nicefrac}       
\usepackage{microtype}      
\usepackage{lipsum}
\usepackage{footnpag}
\usepackage{graphicx}
\usepackage{amsmath}
\usepackage{amssymb}
\usepackage{amsthm}
\usepackage{amsfonts}
\usepackage{algorithm}
\usepackage{multicol}
\usepackage{caption}
\usepackage{subcaption}
\usepackage{rotating}
\usepackage[toc,page]{appendix}
\usepackage{algpseudocode}
\usepackage{blindtext}
\usepackage{textcomp}
\usepackage{gensymb}
\usepackage{rotating}
\graphicspath{ {./} }
\usepackage{biblatex} 
\title{Time Invariant Sensor Tasking for Catalog Maintainance of LEO Space objects using Stochastic Geometry}

\author{
Partha Chowdhury\\PhD Student\\Department of ECE\\ IIIT Delhi\\ \textit{parthac@iiitd.ac.in}
\And
Harsha M\\PhD Student\\Department of ECE\\ IIIT Delhi\\ \textit{harsham@iiitd.ac.in}
\And
Chinni Prabhunath George\\Scientist SG\\ ISRO Telemetry, Tracking and Command Network
\And
Arun Balaji Buduru\\Associate Professor\\Department of CSE\\ IIIT Delhi\\ \textit{arun@iiitd.ac.in}
\And
Sanat K Biswas\\Assistant Professor\\Department of ECE\\ IIIT Delhi\\ \textit{sanat@iiitd.ac.in}
}

\begin{document}
\maketitle
\begin{abstract}
Catalog maintainance of space objects by limited number of ground-based sensors presents a formidable challenging task to the space community. This article presents a methodology for time-invariant tracking and surveillance of space objects in low Earth orbit (LEO) by optimally directing ground sensors. Our methodology aims to maximise the expected number of space objects from a set of ground stations by utilizing concepts from stochastic geometry, particularly the Poisson point process. We have provided a systematic framework to understand visibility patterns and enhance the efficiency of tracking multiple objects simultaneously. Our approach contributes to more informed decision-making in space operations, ultimately supporting efforts to maintain safety and sustainability in LEO. 
 
\end{abstract}


\section{Introduction}
There are currently 10,765 active satellites, most of which are operational in the LEO\cite{noauthor_celestrak_nodate}. In recent years, the rapid increase in satellites and other space objects has heightened the importance of effective tracking systems for space situational awareness. These systems are crucial for ensuring the safety and sustainability of space operations, as collisions between objects can create space debris, posing significant risks to both current and future missions. The challenge lies in the ability to track multiple objects simultaneously with high accuracy using a limited number of ground-based sensors.

Sensor Network management in the context of SSA involves assigning and pointing a set of ground-based sensors for the surveillance and tracking of space objects. Optimal Sensor allocation facilitates maintaining an accurate catalogue and tracking of space objects in LEO \cite{xue2024review}. Optimal sensor tasking for catalogue maintenance can be performed by optimizing various objective functions, primarily focusing on state control and information gain \cite{xue2024review}. In state control for catalogue maintenance, the Posterior Effective Number of Targets (PENT) can be an objective function to be optimized by the sensor tasking algorithm, which works with the recursive estimation frameworks. This objective function calculates the posterior adequate number of space objects by integrating the prior information and sensor observations \cite{mahler2004probabilistic}.

Information gain as an objective function to obtain optimal sensor tasking is another widely used methodology. A method in \cite{adurthi2023dynamic} has been proposed to optimize sensor network parameters by maximizing information gain. The approach utilizes the Fisher Information Matrix (FIM) and Mutual Information (MI) as metrics for information gain. By maximizing an objective function based on these metrics, the method optimizes the pointing direction of sensors for UAVs and satellites. However, these discussed methods are data-intensive and computationally complex. These time-dependent techniques require observations at every epoch and incorporate data fusion to reduce the uncertainty of the solutions to the sensor tasking problem.

Traditional tracking methodologies often struggle with the computational complexity of monitoring the vast number of space objects, especially in dynamic environments where objects frequently change orbits. Our research addresses this issue by introducing a novel methodology for time-invariant tracking of space objects in LEO. By optimally pointing ground sensors, we utilize concepts from stochastic geometry, particularly the Poisson point process, to model the number of space objects visible from a ground station. This model considers factors such as latitude, the inclination of orbits, and the pointing directions of ground sensors.

Modelling of sensor networks using stochastic geometry has been used in various applications, such as the estimation of coverage of satellites for the downlink of wireless communication \cite{lee2022coverage}, coverage analysis \cite{okati2023stochastic}. Stochastic geometry analyzes the spatially averaged performance of any network. Sensor networks are modelled using a novel stochastic geometry framework by developing an Isotropic Satellite Cox Point Process \cite{choi2024cox}. In recent works, Satellite network is analyzed using a Poisson Point Process(PPP) based model \cite{choi2024cox}\cite{al2021analytic}\cite{al2021optimal}, homogeneous binomial point process (BPP) on spherical surfaces\cite{talgat2020nearest}\cite{talgat2020stochastic} \cite{okati2020downlink}. In this work, we have extended this concept to spherical shell encapsulating all LEO objects. 

We have designed an optimization problem to maximize the expected number of space objects observed within the overlapping volume of multiple Ground Sensors. This approach enhances the efficiency of tracking systems and significantly reduces the computational complexity, making it a promising solution for future SSA initiatives.

\section{Problem Formulation}

\begin{figure}[ht]
    \centering
    \includegraphics[width=0.5\linewidth]{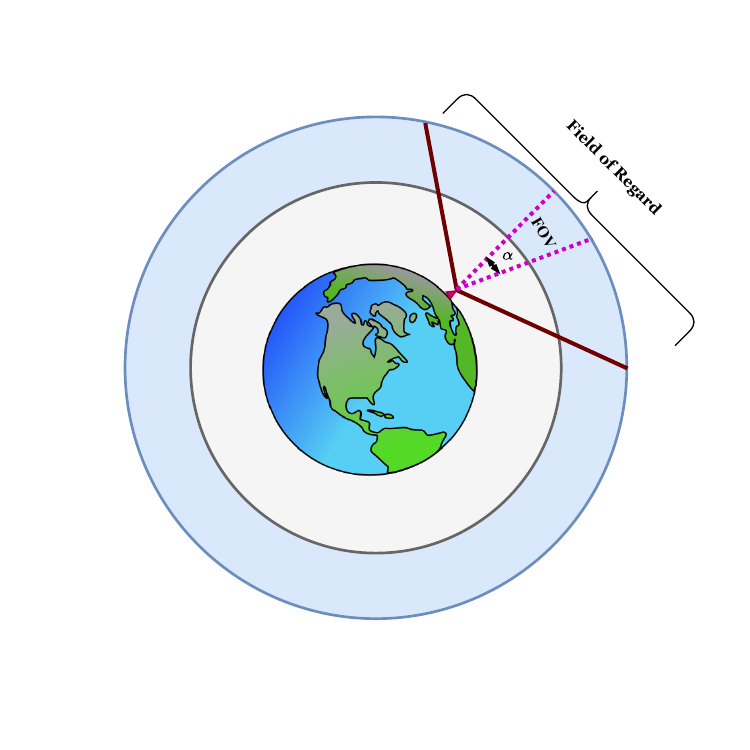}
    \caption{Field of Regard and Field of View}
    \label{fig: FOV- FOR}
\end{figure}

Consider a ground-based sensor network comprising of $M$ number of sensors that are tracking the space objects in the LEO region. Each sensor $i$ of the network has certain Field Of Regard (FOR) and Field Of View (FOV). FOR is the total area that a movable sensor can capture. Similarly, FOV is the angular section, a sensor sees at any given time. The area in the FOV of a sensor $i$ depends upon its FOV angle $\alpha_i$. Figure \ref{fig: FOV- FOR} illustrates the FOV and FOR of the ground-based sensor. Our objective is to maximise the effective number of satellites $\sum_{i=1}^M E[N_i]$ that can be tracked by the sensor network by optimally pointing the sensors towards the space objects.
We are considering $(\phi_i,\lambda_i)$ be the latitude and longitude, and $(E_i, A_i)$ be the pointing directions of each sensor $i$ respectively. Let $R_{min}$ be the minimum altitude of space objects in LEO and $R_{max}$ is the maximum altitude of space objects in LEO. We define the LEO region as the region between two spherical shells of radii $R_{min}$ and $R_{max}$. We calculate the effective latitude $(\phi_e)$ of the observed space objects by the pointing direction of each sensor defined by elevation angle $E_i$ and azimuth angle $A_i$ as:

\begin{equation}
    \phi_e = {\sin^{-1} }\left(\frac{r\cdot \cos{E_i}\cdot \cos{A_i}\cdot \cos{\phi_i}-r\cdot \sin{E_i} \cdot \sin{\phi_i}+{R_e}\cdot \sin{\phi_i}}{R_e + a}\right)
\end{equation}

Where, $r = -R_e\cdot \sin E_i + \sqrt{R_{max}^2-R_e^2\cdot \cos^2 E_i}$ , $R_e$ is the radius of Earth.\\

\begin{figure}[ht]
    \centering
    \includegraphics[width=0.5\linewidth]{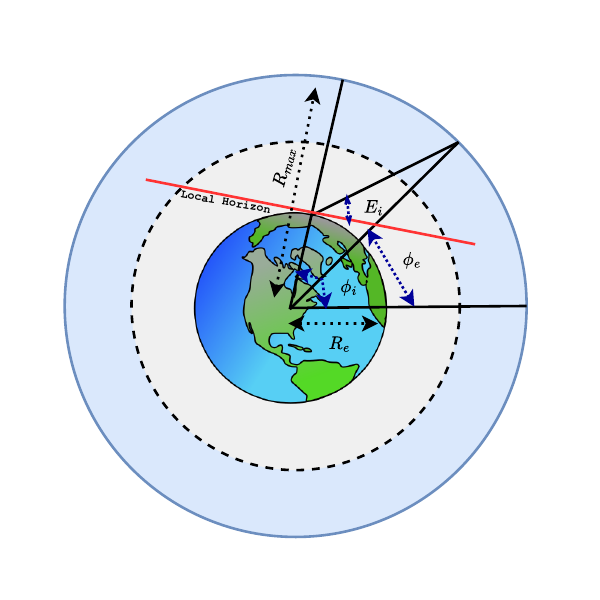}
    \caption{Effective latitude measurement for the observed space objects}
    \label{fig:effective_latitude}
\end{figure}

Figure \ref{fig:effective_latitude} represents the illustration for the effective latitude of the observed space objects at height $R_{max}$. 

For a given number of sensors, the FOV angle of each sensor may vary. In this scenario, we have \( n \) sensors with FOV angles denoted \( \alpha_i \), where \( \alpha_1 > \alpha_2 > \alpha_3 > \ldots > \alpha_n \). This angle $\alpha_i$ covers certain volume $v(\alpha_i)$ of the defined spherical shell, in which visible space objects can be tracked. The effective latitude $\phi_{e}$, the number of visible space objects $N_i$ which are at an inclination $i_i$, and pointing directions $(E_i, A_i)$ are used to compute the satellite density $\delta(\phi_e)$ in a given volume $v(\alpha_{i})$. We consider the distribution of visible space objects within the defined spherical shell as a poisson point process. Therefore, the expected number of visible space objects $E[N_i]$ is the product of volume covered by each sensor $v(\alpha_{i})$ and the space objects density $\delta(\phi_e)$.\\

The objective function for optimizing the pointing direction by maximizing the expected number of space objects is:

\begin{equation}{\label{expected}}
    \mathcal{O}_{N_i} = \sum_{i=1}^M E[N_i]
\end{equation}
The solution for optimized azimuth and elevation angles can be obtained by finding $A_{1 : M}$ and $E_{1 : M}$ which maximizes the objective function as in equation \eqref{eq:opt_sol}

\begin{align}
\label{eq:opt_sol}
    A^*_{1 : M},E^*_{1 : M} &= \underset{A_{1:M},E_{1:M}}{\arg} \max \mathcal{O}_{N_i} \nonumber\\
    & = \underset{A_{1:M},E_{1:M}}{\arg} \max  \sum_{i=1}^M v(\alpha_i) \delta({\phi_i, A_i, E_i})
\end{align}

Where,

\begin{equation}
     \delta(\phi_i,A_i,E_i)=\delta({\phi_e})=\frac{3}{{\sqrt{2}\pi^2}(R_{max}^3 - R_{min}^3)} \sum_{i=0}^k \frac{N_i}{\sqrt{cos(2\phi_e)- cos(2\iota_i)}}
\end{equation}

$N_i$ is the total number of space objects observed in orbit with inclination $\iota_i$, and $k$ is the total number of orbit inclinations and

\begin{align}
    v(\alpha_i)&=\int^{2\pi}_0 \ \int^{(E_i-\frac{\alpha}{2})}_{(E_i+\frac{\alpha}{2})} \ \int^{(R_{max}-R_e)}_{(R_{min}-R_e)} \rho ^2 sin (E_i) \ d\rho \ d(E_i) \ d\phi\nonumber\\
    &= 2\pi \left[\frac{(R_{max}-R_e)^3 - (R_{min}-R_e)^3}{3}\right] \ sin \left(\frac{\alpha}{2}\right) \ sin (E)
\end{align}

Now, the Fisher information gain for the $j^{th}$ space objects measured using $i^{th}$ sensor at $k^{th}$ epoch is
\begin{equation}
    J_j=E[(\nabla_{xj} \ln\Lambda_j(k))\cdot (\nabla_{xj} \ln \Lambda_j(k))']
\end{equation}
Where $\nabla_{xj} \ln \Lambda_j(k) = \sum_{i=1}^n \nabla_{xj}\ln \Lambda_{ij}(k)$, as we consider the observation from each sensor is independent of each other. The likelihood function for observations is given by $\Lambda_{ij} = p(z_{ij}(k)|x_j)$ where $z_{ij}(k)$ is the observations for $j^{th}$ space objects on $i^{th}$ sensor at $k^{th}$ epoch. Consequently, by optimising the objective function $\mathcal{O}$ in equation \eqref{expected}, we also ensure the increase in Fisher information gain.

\section{Solution Approach}

Let us first focus on the overlapping volume of the FOVs of multiple sensors within the LEO. Then, within that area, we maximize the expected number of space objects.

 The FOVs of sensors can overlap only if their respective FORs are overlapping; even if the FORs merely touch each other, we will exclude those sensors from our consideration. We ensure the overlapping of FORs of two sensors in the first step and then ensure the overlapping of FOVs of two sensors with latitude and logitude \((\phi_1, \lambda_1)\) and \((\phi_2, \lambda_2)\), respectively. 

\begin{figure}[ht]
    \centering
    \includegraphics[width=0.5\linewidth trim= 0 180 0 0,clip]{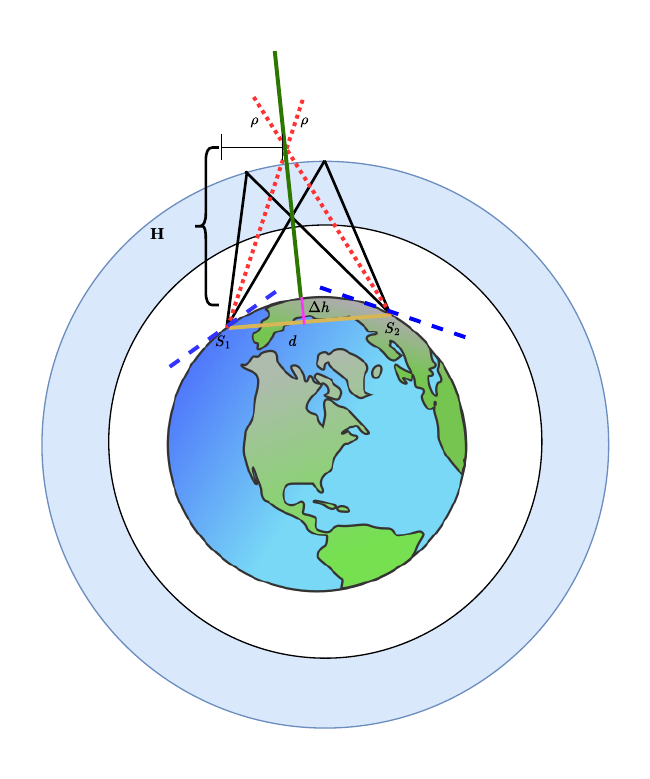}
    \caption{$H$ and $\Delta h$ measurements}
    \label{fig: Intersecting Region}
\end{figure}

Figure \ref{fig: Intersecting Region} represents the scenario of overlapping FOVs of sensors $S_1$ and $S_2$. Assuming a sensor can cover $360^\circ$ on its Azimuth and $90^\circ$, its FOR is a half-hemisphere. Two half hemispheres can overlap if and only if \(2\rho > d\), where \(\rho\) is the maximum range of the sensors, which we consider it as the radius of the hemisphere and d is the eucledian disance. In ECEF reference frame $S_1$ and $S_2$ will be:\\

\begin{multicols}{2}
\begin{equation*}
    S_1 = \begin{bmatrix}
    R_e \ cos (\phi_1) \ cos(\lambda_1)\\
    R_e \ cos (\phi_1) \ sin(\lambda_1)\\
    R_e \ sin (\phi_1)\\
\end{bmatrix}=\begin{bmatrix}
    X_1\\ Y_1\\ Z_1\\
\end{bmatrix}
\end{equation*}

\begin{equation*}
    S_2 = \begin{bmatrix}
    R_e \ cos (\phi_2) \ cos(\lambda_2)\\
    R_e \ cos (\phi_2) \ sin(\lambda_2)\\
    R_e \ sin (\phi_2)\\
\end{bmatrix}=\begin{bmatrix}
    X_2\\ Y_2\\ Z_2\\
\end{bmatrix}
\end{equation*}
\end{multicols}

The Euclidean distance $d$ between $S_1$ and $S_2$ is given by: 
\begin{equation}
    d=\sqrt{(X_2 - X_1)^2 + (Y_2 - Y_1)^2 + (Z_2 - Z_1)^2}
\end{equation}

When considering the overlapping FORs, we must focus on the overlapping area, which should fall within the LEO range, specifically between the minimum height $R_{min}$ and maximum height $R_{max}$ from the centre of the Earth. To ensure this, we set the intersecting point of sensors agree to the condition, such that \(H > (R_{max} - R_e + \Delta h)\). $H$ and $\Delta h$ is shown in Figure \ref{fig: Intersecting Region} only under these conditions can we assert that the maximum point of the overlapping section may lie within the LEO range.\\

H and \(\Delta h\) are expressed as, 
\begin{multicols}{2}
    \begin{equation*}
    H= \sqrt{\rho^2 -\frac{d^2}{4}}
\end{equation*}
\begin{equation*}
    \Delta h = R_e - \sqrt{R_e^2 - \frac{d^2}{4}}
\end{equation*}
\end{multicols}

When considering the lower limit of the LEO as a reference line, ensuring that the imaginary cones formed by the FOV of any two sensors intersect the lower boundary of the LEO at a single point is essential. Therefore, it is necessary to calculate the lower and upper bounds of the elevation angle for each sensor to ensure that they do not exceed the LEO boundary.
Therefore, for each sensor, the lower limit of the elevation angle becomes:

\begin{equation}\label{eq:E-min}
    E_{min}= sin^{-1}\left[\frac{R_{min}^2 - R_e ^2 - \rho^2}{2.R_e.\rho}\right]+\frac{\alpha_i}{2}
\end{equation}

Similarly,

\begin{equation}\label{eq:E-max}
    E_{max}= sin^{-1}\left[\frac{R_{max}^2 - R_e ^2 - \rho^2}{2.R_e.\rho}\right]-\frac{\alpha_i}{2}
\end{equation}

Where $\rho$ is the range of the sensors.

The objective function for calculating the expected number of objects, given the initial elevation and azimuth angles for entire sensor network, is described in algorithm in the appendix \ref{appendix:Algorithm1}. we consider a cluster of sensors that will be overlapping and form a common intersection of volume $v(\alpha_i)$, where $\alpha_i$ represents the smallest FOV angle among the sensors in the overlapping sensor cluster. We have consider $E_{min}$ and $E_{max}$ as the elevation bounds, for maximizing the objective function. 

\section{Sensor Network Data Simulation}

\begin{figure}[ht]
    \centering
    \includegraphics[width=0.46\textwidth]{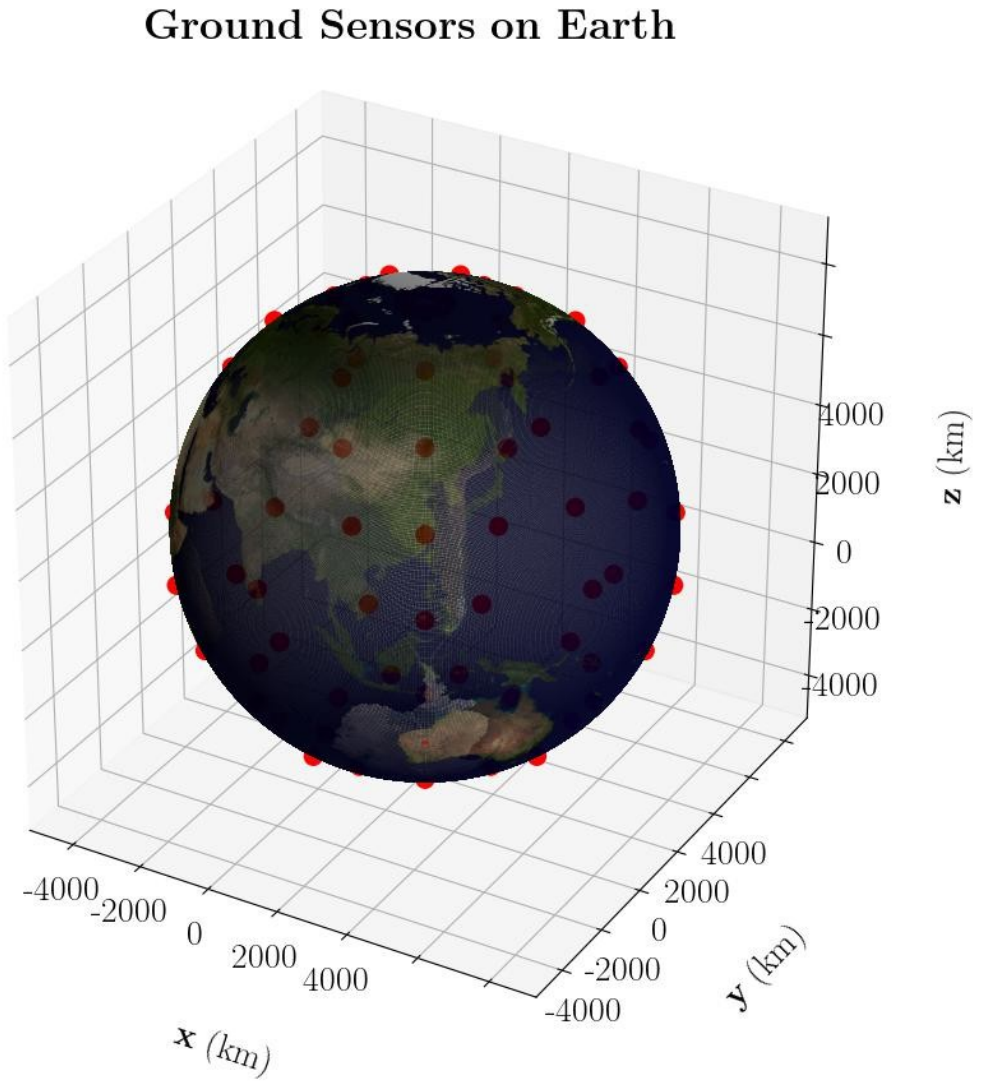}
    \caption{Ground-based sensors visualisation}
    \label{fig:gs_simulation}
\end{figure}

We have considered 100 ground-based sensors for simulation. The latitudes and longitudes are evenly spaced, within the ranges of -85 to 85 degrees and -180 to 180 degrees, respectively. Figure \ref{fig:gs_simulation} shows the placement of sensors on Earth from the identified locations. We have identified 26701 space objects for simulation, and accessed their positions using Two Line Element (TLE) data from \href{www.space-track.org}{\textbf{www.space-track.org}}.  We have recorded the range, azimuth, elevation, and visibility metrics of all space objects from each ground sensor for 360 minutes with 1 min interval in our simulations\footnote[1]{More information on simulation data is in \href{https://github.com/HarshaSSL/JSTOR_AAS}{\textbf{https://github.com/HarshaSSL/JSTOR\_AAS}} }. We have performed these simulations in the server having AMD EPYC 7282 16-Core Processor, 260 GB RAM.  We have used Python 3.8 with the Poliastro \cite{rodriguezPoliastroAstrodynamicsLibrary2016} and Astropy\cite{theastropycollaborationAstropyProjectSustaining2022} libraries. The run-time for simulation was approximately 30 hours.

\section*{Result and Analysis}

In this section, we present a detailed examination of the satellite visibility and sensor network performance based on our simulations and optimization techniques. The analysis begins by comparing the distribution of visible satellites to a Poisson probability distribution to assess its suitability in modeling satellite visibility across different geographic regions. Subsequently, we evaluate the sensor network's ability to maximize the expected number of observed objects by employing advanced optimization techniques, focusing on azimuth and elevation angles for ground sensors. These results provide critical insights into the effectiveness of our modeling approach and optimization strategies in achieving comprehensive coverage of the satellite population.

\begin{figure}[ht]
    \centering
    \begin{subfigure}[b]{0.45\textwidth}
        \includegraphics[width=\textwidth]{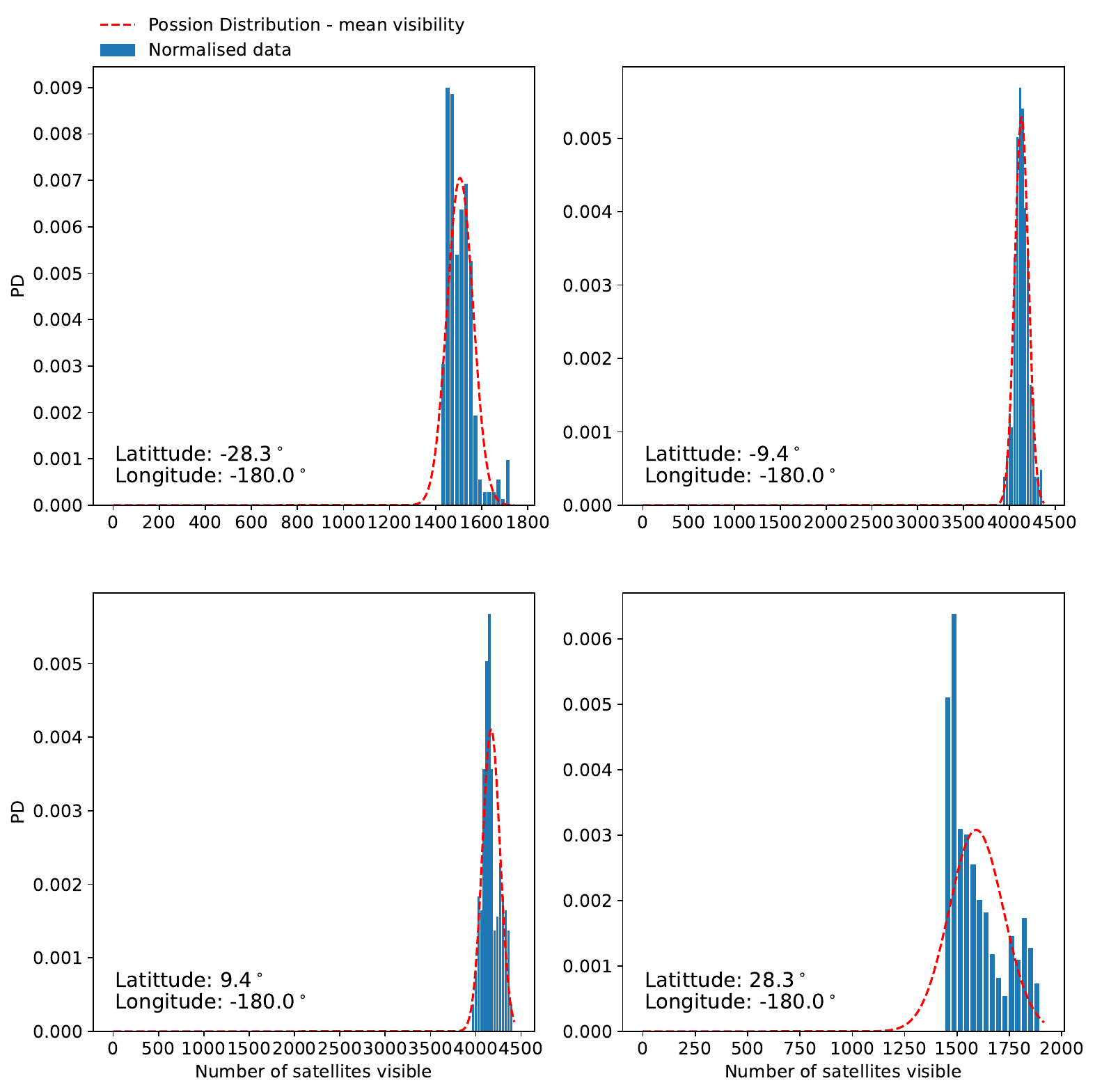}
        \caption{Plot 1}
        \label{fig:poisson_subfig1}
    \end{subfigure}
    \begin{subfigure}[b]{0.45\textwidth}
        \includegraphics[width=\textwidth]{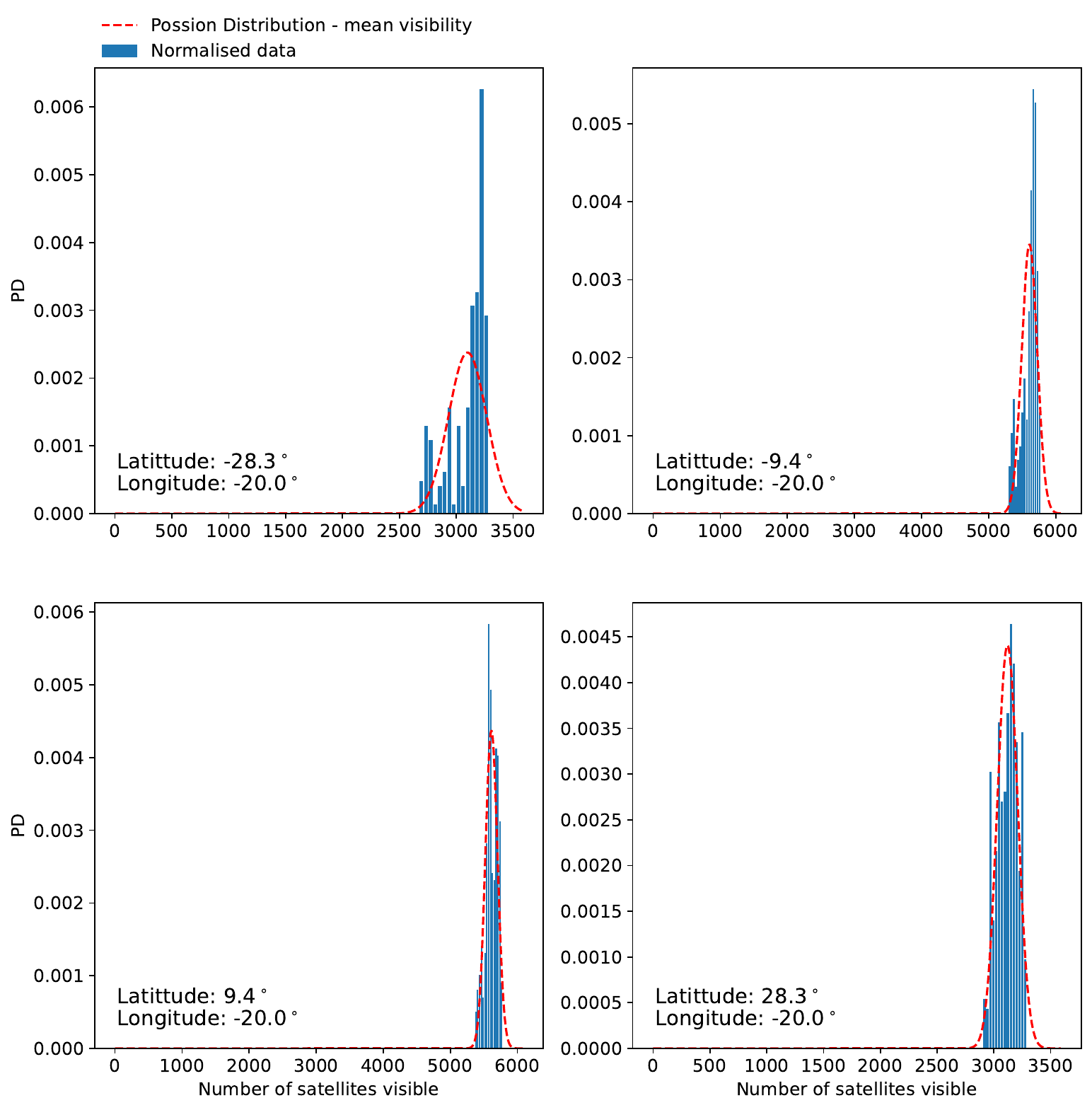}
        \caption{Plot 2}
        \label{fig:poisson_subfig2}
    \end{subfigure}
    \caption{Comparitive analysis of poisson pdf with normalised histogram from visibility data for latitude $-28.3\degree$ to$28.3\degree$ ang for longitude $-180\degree$ and $20\degree$}
    \label{fig:poisson1}
\end{figure}

\begin{figure}[ht]
    \centering
    \begin{subfigure}[b]{0.45\textwidth}
        \includegraphics[width=\textwidth]{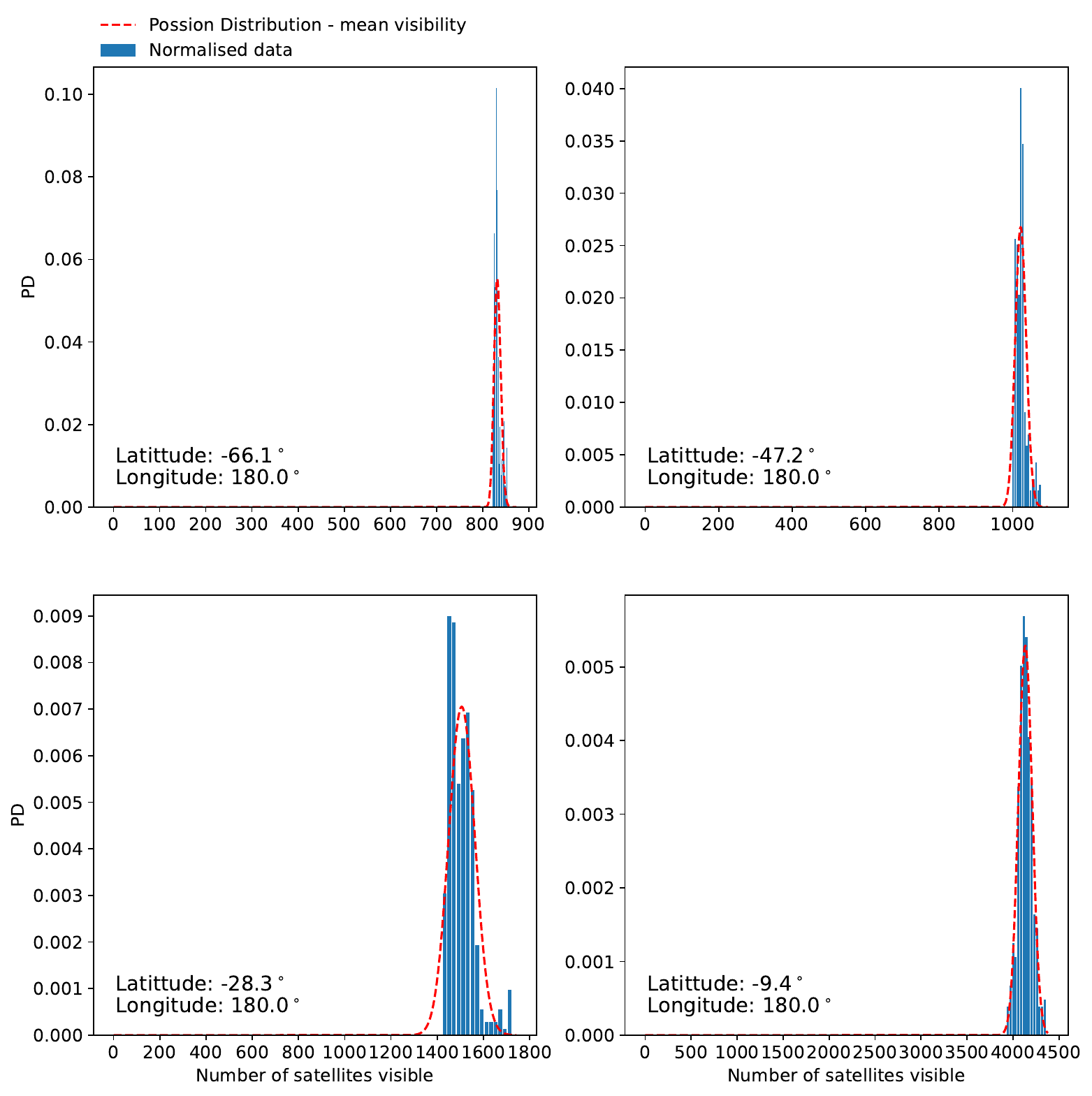}
        \caption{Plot 3}
        \label{fig:poisson_subfig3}
    \end{subfigure}
    \begin{subfigure}[b]{0.45\textwidth}
        \includegraphics[width=\textwidth]{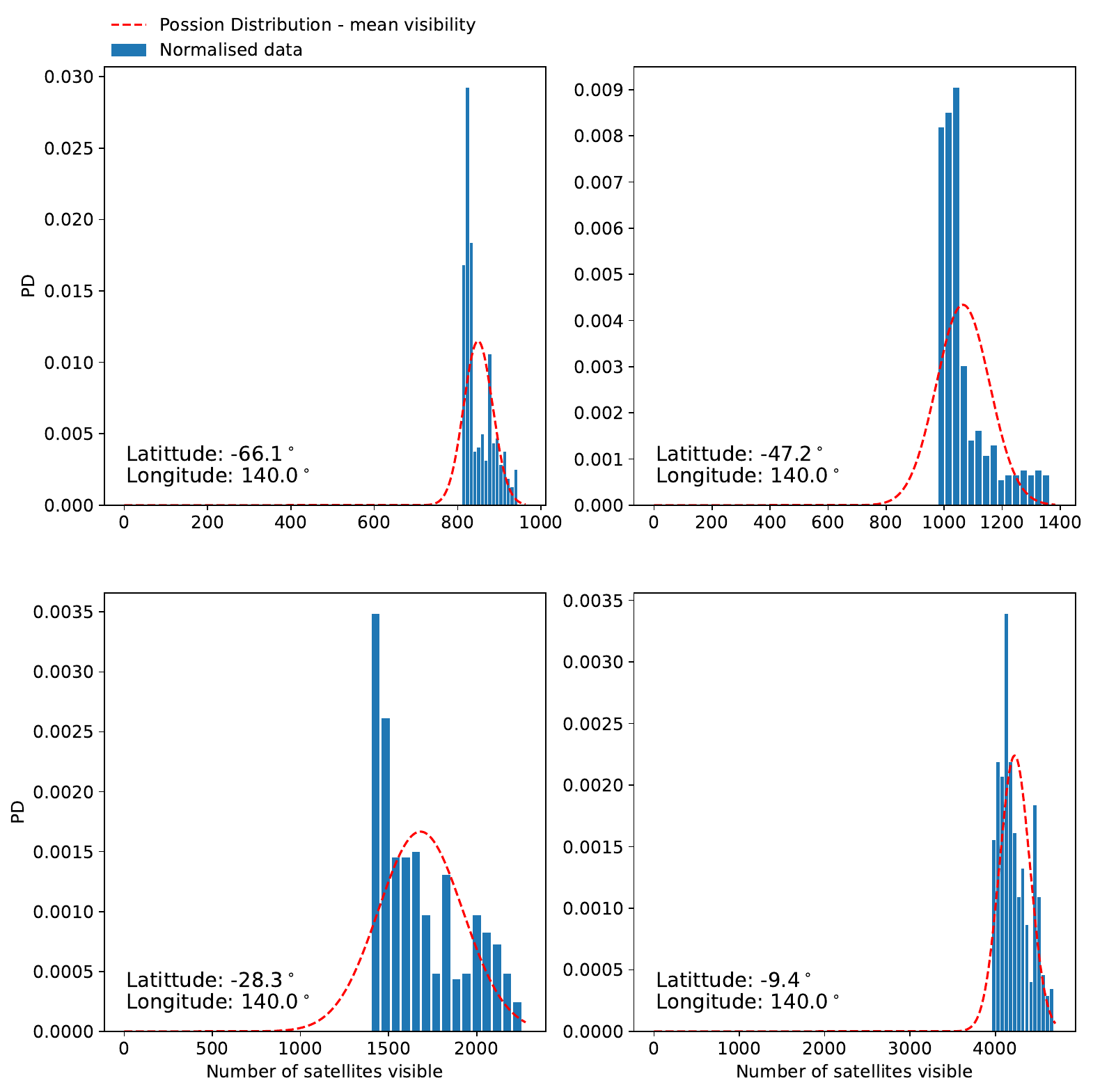}
        \caption{Plot 4}
        \label{fig:poisson_subfig4}
    \end{subfigure}
    \caption{Comparitive analysis of poisson pdf with normalised histogram from visibility data latitude $-66.1\degree$ to$-9.4\degree$ and for longitude $180\degree$ and $140\degree$}
    \label{fig:poisson2}
\end{figure}

Figure \ref{fig:poisson1} and Figure \ref{fig:poisson2} illustrates the probability density of the effective number of satellites visible from simulated ground sensor locations across various latitudes and longitudes. Specifically, Figure \ref{fig:poisson_subfig1} depicts the probability density for ground sensors at latitudes between $-28.3\degree$ and $28.3\degree$ at longitude $-180\degree$, while Figure \ref{fig:poisson_subfig2} shows the same for longitude $-20\degree$. Similarly, Figure \ref{fig:poisson_subfig3} represents latitudes from $-9.4\degree$ to $-66\degree$ at longitude $180\degree$, and Figure \ref{fig:poisson_subfig4} corresponds to latitudes from $-66\degree$ to $-9.4\degree$ at longitude $140\degree$. 
These figures illustrate a comparison between the Poisson probability distribution and the normalized histogram derived from the satellite visibility data, which we simulated from the simulation data. \footnote[2]{More information on simulation data is in \href{https://github.com/HarshaSSL/JSTOR_AAS}{\textbf{https://github.com/HarshaSSL/JSTOR\_AAS}} }. While they do not conclusively prove our hypothesis, they demonstrate that the Poisson probability distribution function closely approximates the normalized distribution of visible satellites when effective latitude is used as a distribution parameter. 

By evaluating the sensor network to find the overlapping volumes, using the expressions $H$ and $\delta h$, it resulted in 10 unique overlapping volumes. These volumes are covered using 30 sensors for 8 volumes and 20 sensors for 2 volumes, respectively. To maximise the expected number of objects in the objective function, we have applied \textbf{``trust-constr''} (A trust-region algorithm designed for constrained optimisation problems) optimisation method from SciPy library \cite{2020SciPy-NMeth} and \textbf{``Particle Swarm Optimisation''} method using pyswarm library \cite{tisimst_tisimstpyswarm_2025} for finding the optimal azimuth and elevation angles for each sensor. We have set the bounds for elevation $E_{min}$ as $0\degree$ and $E_{max}$ as $90\degree$ and for azimuth, we have set the bounds $A_{min}$ as $0\degree$ and $A_{max}$ as $360\degree$. We initialised both Azimuth and Elevation angles for each sensor as $10\degree$. 

\begin{figure}[ht]
\centering
\subfloat[Results of Trust-constraint Optimisation]{\includegraphics[width=0.45\textwidth, keepaspectratio]{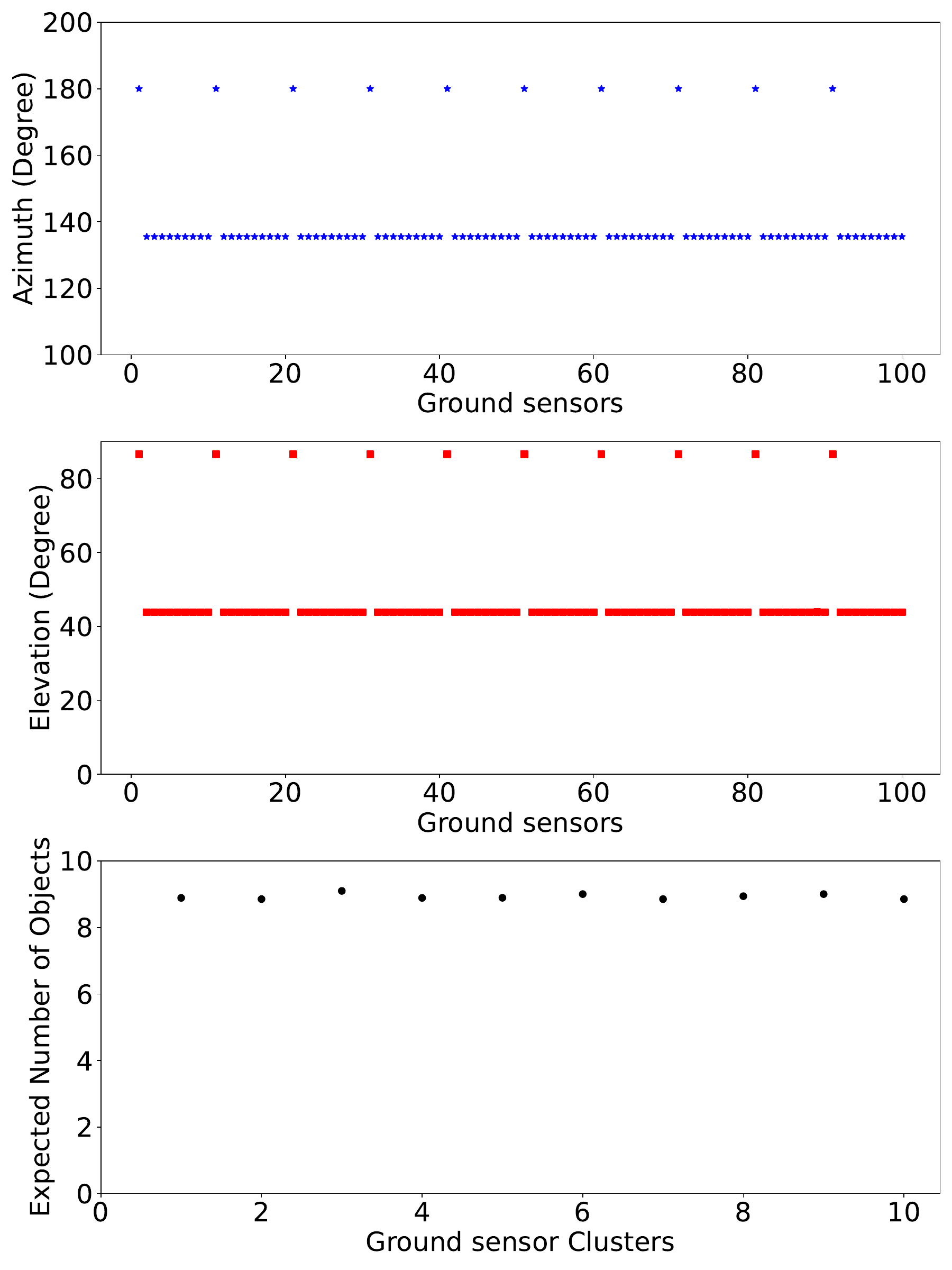}\label{fig:optimiser_trust}}\hfill
\subfloat[Results of Particle Swarm Optimisation]{\includegraphics[width=0.45\textwidth, keepaspectratio]{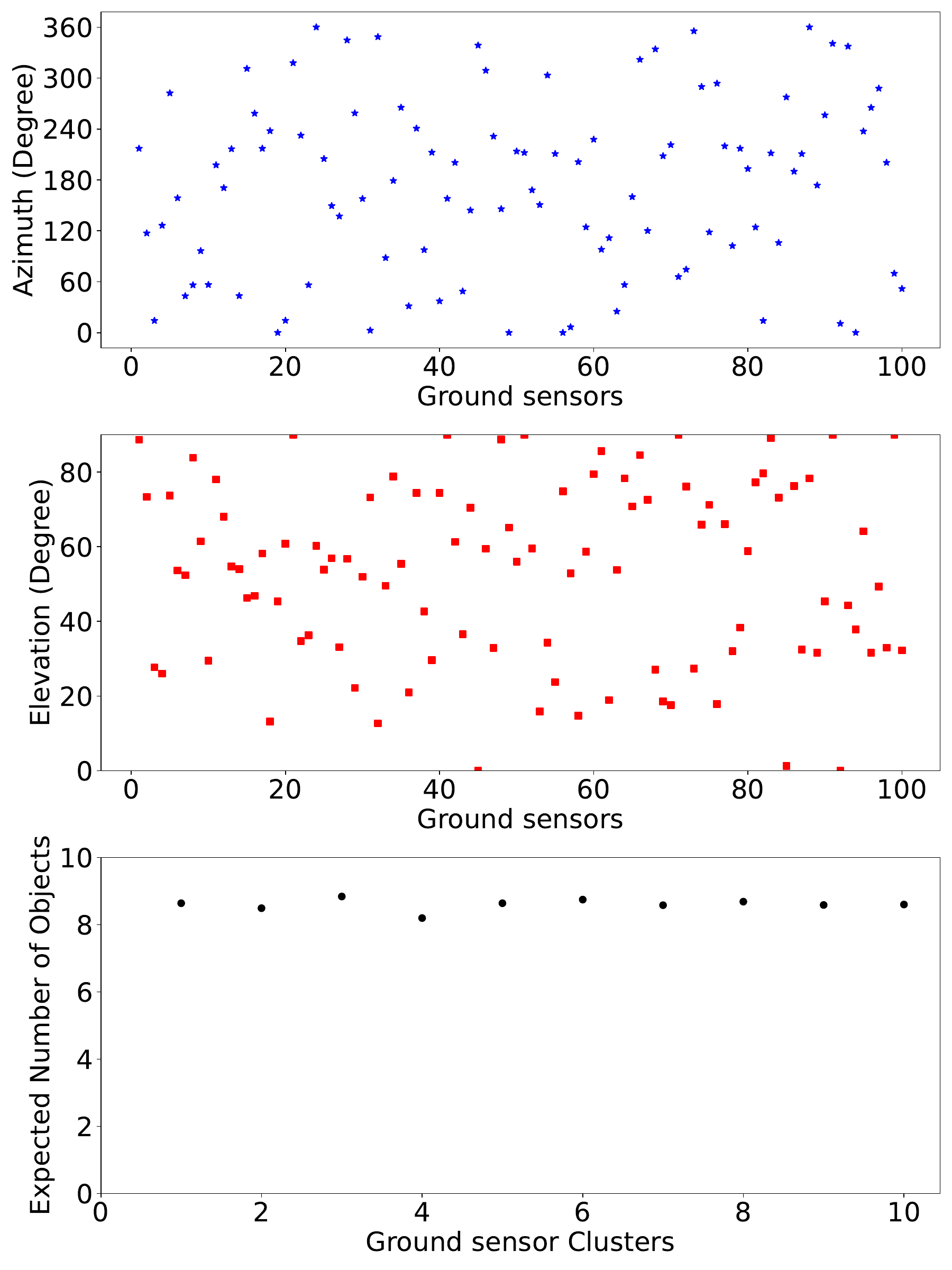}\label{fig:optimiser_pso}}
\caption[Optional caption for list of figures 5-8]{Results of optimisation methods}
\label{fig:optimisation}
\end{figure}

The optimised azimuth and elevation angles and an expected number of space objects visible are plotted against each ground station in Figure \ref{fig:optimisation}. The ``trust-constr'' optimisation method produced 89.2769, and the ``Particle Swarm Optimisation'' method produced 86.4633 as the maximum number of expected space objects. These results are very close to 91.0333 and 89.6222, the number of space objects we measured while verifying the optimised azimuth and elevations deduced from both the methods with real-time dataset respectively. However, in Figure \ref{fig:optimisation}, it is visible that the distribution of optimised azimuth and elevation are different in both methods. This is possible because two different methods may take different maxima values. Further methodological investigations are needed to explore this issue.

\section{Conclusion}

In this article, we have designed a method for time-invariant tracking of space objects in LEO by optimally pointing ground sensors. Our approach leverages concepts from stochastic geometry, particularly the Poisson point process, to effectively model the number of space objects visible from a ground station.

To enhance the effectiveness of our tracking system, we have formulated an optimization problem that may be convex in nature, depending upon certain restrictions. Our main goal was to maximize the total expected number of space objects observed in a particular region where the FOV multiple sensors overlap. 

Our methodology streamlines the process of tracking space objects and significantly reduces computational complexity, making it more feasible to implement in real-world applications. We believe our approach can pave the way for more efficient space situational awareness systems by minimizing the resources needed for computation while maintaining high observation quality.

Moreover, this methodology opens avenues for future research, including the potential integration of machine learning techniques to refine the prediction models further and enhance observation optimization. It also raises interesting questions about the operational parameters necessary for different types of ground sensors and how they can be coordinated to achieve maximum coverage. 

 As the population of satellites and debris in LEO continues to grow, the ability to track space objects effectively is critical for preventing potential collisions and ensuring a stable space environment. By offering a robust framework for monitoring space traffic with reduced computational demands, this work contributes to developing advanced space situational awareness systems.

\section*{Acknowledgement}
This research is funded by the ISRO Respond program, project ref. no. ISRO/RES/3/921 /22-23



\begin{appendices}
\section{Appendix 1: Algorithm }

\begin{algorithm}[ht]
\begin{algorithmic}[1]
\Require Positions of sensors $[\alpha_i, \lambda_i]$, range $[\rho]$, Earth radius $[R_e]$, LEO bounds $[R_{\text{max}]}$ and $[R_{\text{min}}]$, sensor angles $[\alpha]$

\Ensure Number of sensors $n\geq 2$

\State Initialize $V = 0$\label{appendix:Algorithm1}
\For{each sensor pair $i$ and $j$ where $i \neq j$}
    \State Compute distance $d \gets \text{Euclidean Distance}(\textit{Sensor Positions}[i], \textit{Sensor Positions}[j])$
    \If{$2 \rho > d$ and $H > (R_{max}-R_e + \Delta h)$}
        \State Make cluster of sensors for a $i$ which k sensors are overlapping.
        \State If both satisfy, compute lower  elevation angle of  each sensor j in the cluster:
        \[
        E_{\text{min}_j} = \sin^{-1} \left( \frac{R_{\text{min}}^2 - R_e^2 - \rho^2}{2 R_e \rho} \right) + \frac{\alpha_j}{2}\] 
        \State Compute upper elevation angle of each sensor j in the cluster:
        \[E_{\text{max}_j}  = \sin^{-1} \left( \frac{R_{\text{max}}^2 - R_e^2 - \rho^2}{2 R_e \rho} \right) - \frac{\alpha_j}{2}\]
        
        \State Compute overlapping volume:
        \[
        v(\alpha_i) = 2\pi \left[\frac{(R_{max}-R_e)^3 - (R_{min}-R_e)^3}{3}\right] \ sin \left(\frac{\alpha_i}{2}\right) \ sin (E_i)
        \]
        \State Compute \[\delta(\phi_i,A_i,E_i)=\delta({\phi_e})=\frac{3}{{\sqrt{2}\pi^2}(R_{max}^3 - R_{min}^3)} \sum_{i=0}^k \frac{N_i}{\sqrt{cos(2\phi_e)- cos(2\iota_i)}}\]
         \EndIf
         \State Calculate minimum of $v(\alpha_i)$
        \State Apply $E_{min_i}$ \& $E_{max_i}$ as bounds for solving optimisation problem
    \EndFor
        \State $A^*_{1 : M},E^*_{1 : M} = \underset{A_{1:M},E_{1:M}}{\arg} \max \  \mathcal{O}_{N_i} = \underset{A_{1:M},E_{1:M}}{\arg} \max \  \sum_{i=1}^M v(\alpha_i) \delta({\phi_i, A_i, E_i})$

\Return $\mathcal{O}_{N_i}$
\end{algorithmic}
\end{algorithm}
\end{appendices}

\printbibliography

@article{okati2023stochastic,
  title={Stochastic Coverage Analysis for Multi-Altitude LEO Satellite Networks},
  author={Okati, Niloofar and Riihonen, Taneli},
  journal={IEEE Communications Letters},
  year={2023},
  publisher={IEEE}
}

@inproceedings{mahler2004probabilistic,
  title={Probabilistic objective functions for sensor management},
  author={Mahler, Ronald PS and Zajic, Tim R},
  booktitle={Signal Processing, Sensor Fusion, and Target Recognition XIII},
  volume={5429},
  pages={233--244},
  year={2004},
  organization={SPIE}
}

@article{xue2024review,
  title={Review of sensor tasking methods in Space Situational Awareness},
  author={Xue, Chenbao and Cai, Han and Gehly, Steve and Jah, Moriba and Zhang, Jingrui},
  journal={Progress in Aerospace Sciences},
  volume={147},
  pages={101017},
  year={2024},
  publisher={Elsevier}
}

@incollection{adurthi2023dynamic,
  title={Dynamic Data-Driven Sensor Tasking with Applications in Space and Aerospace Systems},
  author={Adurthi, Nagavenkat and Singla, Puneet and Majji, Manoranjan},
  booktitle={Handbook of Dynamic Data Driven Applications Systems: Volume 2},
  pages={249--283},
  year={2023},
  publisher={Springer}
}

@article{lee2022coverage,
  title={Coverage analysis of LEO satellite downlink networks: Orbit geometry dependent approach},
  author={Lee, Junse and Noh, Song and Jeong, Sooyeob and Lee, Namyoon},
  journal={arXiv preprint arXiv:2206.09382},
  year={2022}
}

@article{choi2024cox,
  title={Cox Point Processes for Multi Altitude LEO Satellite Networks},
  author={Choi, Chang-Sik and others},
  journal={IEEE Transactions on Vehicular Technology},
  year={2024},
  publisher={IEEE}
}

@article{talgat2020nearest,
  title={Nearest neighbor and contact distance distribution for binomial point process on spherical surfaces},
  author={Talgat, Anna and Kishk, Mustafa A and Alouini, Mohamed-Slim},
  journal={IEEE Communications Letters},
  volume={24},
  number={12},
  pages={2659--2663},
  year={2020},
  publisher={IEEE}
}

@article{talgat2020stochastic,
  title={Stochastic geometry-based analysis of LEO satellite communication systems},
  author={Talgat, Anna and Kishk, Mustafa A and Alouini, Mohamed-Slim},
  journal={IEEE Communications Letters},
  volume={25},
  number={8},
  pages={2458--2462},
  year={2020},
  publisher={IEEE}
}

@article{okati2020downlink,
  title={Downlink coverage and rate analysis of low earth orbit satellite constellations using stochastic geometry},
  author={Okati, Niloofar and Riihonen, Taneli and Korpi, Dani and Angervuori, Ilari and Wichman, Risto},
  journal={IEEE Transactions on Communications},
  volume={68},
  number={8},
  pages={5120--5134},
  year={2020},
  publisher={IEEE}
}

@article{al2021analytic,
  title={An analytic approach for modeling the coverage performance of dense satellite networks},
  author={Al-Hourani, Akram},
  journal={IEEE Wireless Communications Letters},
  volume={10},
  number={4},
  pages={897--901},
  year={2021},
  publisher={IEEE}
}

@article{al2021optimal,
  title={Optimal satellite constellation altitude for maximal coverage},
  author={Al-Hourani, Akram},
  journal={IEEE Wireless Communications Letters},
  volume={10},
  number={7},
  pages={1444--1448},
  year={2021},
  publisher={IEEE}
}

@inproceedings{rodriguezPoliastroAstrodynamicsLibrary2016,
  title = {Poliastro: An Astrodynamics Library Written in Python with Fortran Performance},
  booktitle = {6th {{International Conference}} on {{Astrodynamics Tools}} and {{Techniques}}},
  author = {Rodr{\'i}guez, Juan Luis Cano and Eichhorn, Helge and McLean, Frazer},
  year = {2016}
}

@misc{theastropycollaborationAstropyProjectSustaining2022,
  title = {The {{Astropy Project}}: {{Sustaining}} and {{Growing}} a {{Community-oriented Open-source Project}} and the {{Latest Major Release}} (v5.0) of the {{Core Package}}},
  shorttitle = {The {{Astropy Project}}},
  author = {{Price-Whelan}, Adrian M and Sip{\H o}cz, {\relax BM} and G{\"u}nther, {\relax HM} and Lim, {\relax PL} and Crawford, {\relax SM} and Conseil, S and Shupe, {\relax DL} and Craig, {\relax MW} and Dencheva, N and Ginsburg, A and others},
  year = {2022},
  month = jun,
  eprint = {2206.14220},
  primaryclass = {astro-ph},
  doi = {10.3847/1538-4357/ac7c74},
  urldate = {2023-02-15},
  archiveprefix = {arxiv}
}

@misc{noauthor_celestrak_nodate,
	title = {{CelesTrak}: {Active} {Satellites}},
	url = {https://celestrak.org/NORAD/elements/table.php?GROUP=active&FORMAT=tle},
	urldate = {2025-01-08},
}

@ARTICLE{2020SciPy-NMeth,
  author  = {Virtanen, Pauli and Gommers, Ralf and Oliphant, Travis E. and
            Haberland, Matt and Reddy, Tyler and Cournapeau, David and
            Burovski, Evgeni and Peterson, Pearu and Weckesser, Warren and
            Bright, Jonathan and {van der Walt}, St{\'e}fan J. and
            Brett, Matthew and Wilson, Joshua and Millman, K. Jarrod and
            Mayorov, Nikolay and Nelson, Andrew R. J. and Jones, Eric and
            Kern, Robert and Larson, Eric and Carey, C J and
            Polat, {\.I}lhan and Feng, Yu and Moore, Eric W. and
            {VanderPlas}, Jake and Laxalde, Denis and Perktold, Josef and
            Cimrman, Robert and Henriksen, Ian and Quintero, E. A. and
            Harris, Charles R. and Archibald, Anne M. and
            Ribeiro, Ant{\^o}nio H. and Pedregosa, Fabian and
            {van Mulbregt}, Paul and {SciPy 1.0 Contributors}},
  title   = {{{SciPy} 1.0: Fundamental Algorithms for Scientific
            Computing in Python}},
  journal = {Nature Methods},
  year    = {2020},
  volume  = {17},
  pages   = {261--272},
  adsurl  = {https://rdcu.be/b08Wh},
  doi     = {10.1038/s41592-019-0686-2},
}

@misc{tisimst_tisimstpyswarm_2025,
	title = {tisimst/pyswarm},
	url = {https://github.com/tisimst/pyswarm},
	urldate = {2025-01-08},
	author = {tisimst},
	month = jan,
	year = {2025},
	note = {original-date: 2014-02-22T14:29:59Z},
}

\footnote{This paper has been accepted and presented at the 35th AIAA/AAS Space Flight Mechanics Meeting, Kaua'i, Hawaii. Paper no.: AAS 25-325}
\end{document}